\documentclass[onecolumn,amsmath,amssymb,floatfix,aps,superscriptaddress]{revtex4}
\usepackage{subfigure}
\usepackage{color}

\usepackage{graphicx}
\usepackage{bm}  
\usepackage{epsfig}
\usepackage{mathtools}
\usepackage{verbatim} 

\begin{document}

\title{A generalized conservation law for main-chain polymer nematics}

\author{Daniel Sven\v sek}
\affiliation{Department of Physics, Faculty of Mathematics and Physics, University of Ljubljana, Jadranska 19, SI-1000 Ljubljana, Slovenia}

\author{Rudolf Podgornik}
\affiliation{Department of Physics, Faculty of Mathematics and Physics, University of Ljubljana, Jadranska 19, SI-1000 Ljubljana, Slovenia}
\affiliation{Department of Theoretical Physics, J. Stefan Institute, Jamova 39,  SI-1000 Ljubljana, Slovenia}
\affiliation{Department of Physics,  University of Massachusetts, Amherst MA 01003, USA}

\begin{abstract}
We explore the implications of the conservation law(s) and the corresponding ``continuity equation(s)",  resulting from the coupling between the positional and the orientational order in main-chain polymer nematics, by showing that the vectorial and tensorial forms of these equations are in general {\sl not equivalent} and can not be reduced to one another, but neither are they disjoint alternatives. We analyze the relation between them and elucidate the fundamental role that the chain backfolding plays in the determination of their relative strength and importance.  Finally, we show that the correct penalty potential in the effective free energy, implementing these conservation laws, should actually connect both the tensorial and the vectorial constraints. We show that the consequences of the polymer chains connectivity for their consistent mesoscopic description are thus not only highly non-trivial but that its proper implementation is absolutely crucial for a consistent coarse grained description of the main-chain polymer nematics. 
\end{abstract}


\maketitle

\section{Introduction}
\label{sec:intro}

\noindent
Liquid crystalline order is ubiquitous in biological materials \cite{Rey1} and many properties of these systems can be analyzed in terms of the standard Landau-de Gennes approach \cite{Selinger-book}, without particularly worrying about e.g. the polymer nature of the main-chain polymer nematogens. Nevertheless, it was recognized a while ago, that the Landau-de Gennes approach needs to be modified specifically to take into account the polymer nature, i.e. the microscopic connectivity of the underlying mesogens \cite{degennes,meyer}. This connectivity leads to a coupling between the positional and orientational order of the polymer molecules. The ensuing ``continuity equation" derived to take into account this specificity of the polymer mesophase materials was shown to matter fundamentally for a consistent description of macroscopic properties of these systems \cite{selinger,nelson0, nelson,selingerJP}.  In addition, recently we showed that depending on the nature and the symmetry of the mesophase order, this ``continuity equation" might not be the only condition that the coarse grained description of polymer nematics needs to satisfy. In fact, its generalized form in terms of a vectorial conservation law \cite{svensek-podgornik,svensek} suffices for polar nematic order, while in the more usual quadrupolar nematic order case a different, tensorial conservation law has to be included in the Landau-de Gennes description \cite{svensek,errata}. 

Discovering a new form of the conservation law, stemming from the microscopic connectivity of the constitutive nematogen molecules, naturally leads to questions, such as which form of the conservation law is the correct one, how are they related,  when is the use of one preferred to the other one and how important is their implementation for the proper description of polymer nematics. In this note we address and resolve all these fundamental questions for future applications. In fact, we show that the two conservation laws are not equivalent, are irreducible and in general lead to different consequences specifically stemming from the presence of the backfolding configurations of the polymer chains, or in an extreme case, to the presence of localized hairpins and/or kinks which should be in particular relevant in the description of confined DNA mesophases, where such local defects or elastic energy non-linearities have been implicated recently \cite{Hoang}.

In what follows, extending our previous analysis \cite{svensek-podgornik,svensek}, we thus propose a new, generalized conservation law for the main-chain polymer nematics that {\sl consistently incorporates} both the vectorial as well as the tensorial conservation laws. Moreover, and in particular, this includes also a well-grounded interpretation of the recently derived tensorial conservation law that has been so far elusive. We also indicate to what extent and in which cases applying the vectorial constraint to nematic director of the main-chain polymer is invalidated, even in the case of perfectly rigid chains, and leads to fundamentally wrong conclusions. In addition we show that the two constituent conservation laws are manifest in the two extreme cases of flexible and inflexible chains, respectively. The present analysis is also relevant to ascertain the detectability of the two constraints, whether to pinpoint the observable differences between them or the novel features of their joint implementation, for which a detailed simulation approach, on which we recently embarked, seems at present to be our best guide \cite{svensek-correlations}.

\section{Vectorial and tensorial conservation laws}
\label{sec:vecten}

On the macroscopic level, the microscopic connectivity of monomers into the chain of a main-chain polymer nematic shows up as a constraint linking director deformations and density variations. Since its formulation, the accepted form of this constraint has been \cite{degennes,meyer,selinger,nelson0,nelson,selingerJP}
\begin{equation}
	\nabla\cdot(\rho_s {\bf n}) = \rho^+ - \rho^-,
	\label{vectorial-n}
\end{equation}
where $\rho_s({\bf r})$ is the surface density of polymer chains perforating the plane perpendicular to the nematic director ${\bf n}({\bf r})$, while $\rho^+({\bf r})$ and $\rho^-({\bf r})$ are volume densities of the beginnings and endings of chains, respectively.
This constraint has been used in connection with the nematic director, i.e., the principal axis of the nematic order tensor $\sf Q$, describing quadrupolar orientational ordering \cite{degennes-book}.

Including the modulus of the orientational ordering, it becomes apparent that the constraint Eq.~(\ref{vectorial-n}) naturally augments to a vectorial law, i.e., the conservation law for the polymer current density ${\bf j} = \rho\ell_0{\bf a}$ \cite{svensek-podgornik,svensek},
\begin{equation}
	\nabla\cdot(\rho\ell_0{\bf a}) = \rho^+ - \rho^-,
	\label{vectorial}
\end{equation}
where $\rho({\bf r})$ is the volume density of arbitrary polymer segments (e.g. monomers) of length $\ell_0$ and ${\bf a}({\bf r}) = \langle {\bf t}_\alpha\rangle$ is the mesoscopic average of monomer tangents ${\bf t}_\alpha$. In passing, by comparison of Eqs.~(\ref{vectorial-n}) and (\ref{vectorial}) we see that $\rho_s = \rho\ell_0 a$, which can be also straightforwardly inferred from geometry.
A formal derivation \cite{svensek} of Eq.~(\ref{vectorial}) leaves no doubt that this is indeed the exact conservation law for the polar orientational order $\bf a$ of the chain tangents.

However, recently we derived a different,  tensorial conservation law \cite{svensek,errata} for main-chain polymer nematics, of the form
\begin{equation}
	\partial_i\partial_j \tilde{J}_{ij}=\partial_i\partial_j \left[\rho\ell_0\left(Q_{ij}+\textstyle{1\over 2}\delta_{ij}\right)\right] = 
		\textstyle{3\over 2} \partial_i \left(g_i^+ - g_i^-\right),
	\label{tensorial}
\end{equation}
where ${\bf g}({\bf r}) \equiv {\bf g}^+({\bf r}) - {\bf g}^-({\bf r})$ is the volume density of chain head/tail tangents. Conforming to the quadrupolar symmetry, heads and tails are indistinguishable in $\bf g$, with the tangents always pointing away from the head/tail. This purely geometric (kinematic) continuity condition, just like the vectorial analogue Eq.~(\ref{vectorial}), is a consequence of the fact that in a system of unbreakable chains the orientation and the position of a monomer are not independent, and must be satisfied by any configuration ${\sf Q}({\bf r})$.

For a configuration ${\sf Q}({\bf r})$, the requirement Eq.~(\ref{tensorial}) is however not the complete story --- it is a necessary but, in general, not the only condition the tensorial configuration must satisfy.  Namely, since quadrupolar order is insensitive to chain backfolding, the tensorial constraint Eq.~(\ref{tensorial}) is not at all affected by the flexibility of the chain. Therefore, in order to describe the chain rigidity (semi-flexibility), the tensorial constraint requires the backup of the tighter vectorial constraint Eq.~(\ref{vectorial}), as we will propose in what follows. Indeed we will show that inflexible chains (no chain backfolding) should follow closely the vectorial conservation Eq.~(\ref{vectorial}) for the nematic director (which can be defined as a vector in this limit), and that  flexible chains with strong backfolding, and therefore necessarily with a vanishing local mesoscopic polar order of chain tangents, should satisfy the tensorial conservation Eq.~(\ref{tensorial}). The aim here is thus to shed light on the connection between the vectorial and the tensorial conservation laws and to elaborate on the correct application of both conservation laws to the general case of semi-flexible chains, which should include the limits of inflexible and flexible chains as special cases.

\section{Vectorial conservation law for the nematic director}
\label{sec:recovered_polar}

\noindent
The vectorial conservation law has been standardly used for the nematic director, and vice versa, the constraint linking density variations and nematic director deformations has been implemented exclusively in terms of the vectorial conservation law.
However, in principle, this is problematic. The vectorial conservation law is the continuity equation for the polymer current density, which is a vectorial quantity. As such, strictly speaking it cannot be expressed in terms of a quantity with quadrupolar symmetry like the nematic director.
In this Section we present the underlying argument that nevertheless enables the use of the vectorial conservation law also for the nonpolar nematic director. The basis of this argument is the interpretation of chain hairpins (point-like $\sf U$ turns) as effective chain ends. This viewpoint dates back to Odijk \cite{odijk88} and is also reflected in subsequent stat-mech results \cite{terentjev,kamien}. Here we articulate it for general backfoldings, in a way that rigorously connects with the macroscopic vectorial conservation law Eq.~(\ref{vectorial}).

In a system of chains that does not exhibit macroscopic polar order (the chains themselves can be either apolar or polar), one cannot define a vectorial order parameter. If a polar orientational order is nevertheless defined in a local mesoscopic volume by an arbitrary convention of the chain directions (e.g. such as to maximize the so-defined local polar order), it will decay away from this point with a finite characteristic length (represented by the global persistence length \cite{odijk88}) depending on the density of hairpins or the extent of chain backfolding in general. This decay length is not necessarily small and may be comparable to other length scales in the system including the system size. It distinguishes between ensembles of stiff vs. flexible chains and is therefore a physical reality --- not in terms of orientational order (which for apolar phases and, by definition, for apolar chains is insensitive to the extent of backfolding), but in terms of the chain backfolding configurational degrees of freedom.

Following the arguments presented by Odijk, the density of top and bottom hairpins \cite{odijk88} should enter the vectorial conservation law as a source term, similar to the density of chain heads and tails in Eq.~(\ref{vectorial}). But how can one consistently apply the vectorial conservation law if the polar orientational order cannot be defined?

\begin{figure}[h]
\begin{center}
	\mbox{
		\subfigure[]{\includegraphics[width=35mm]{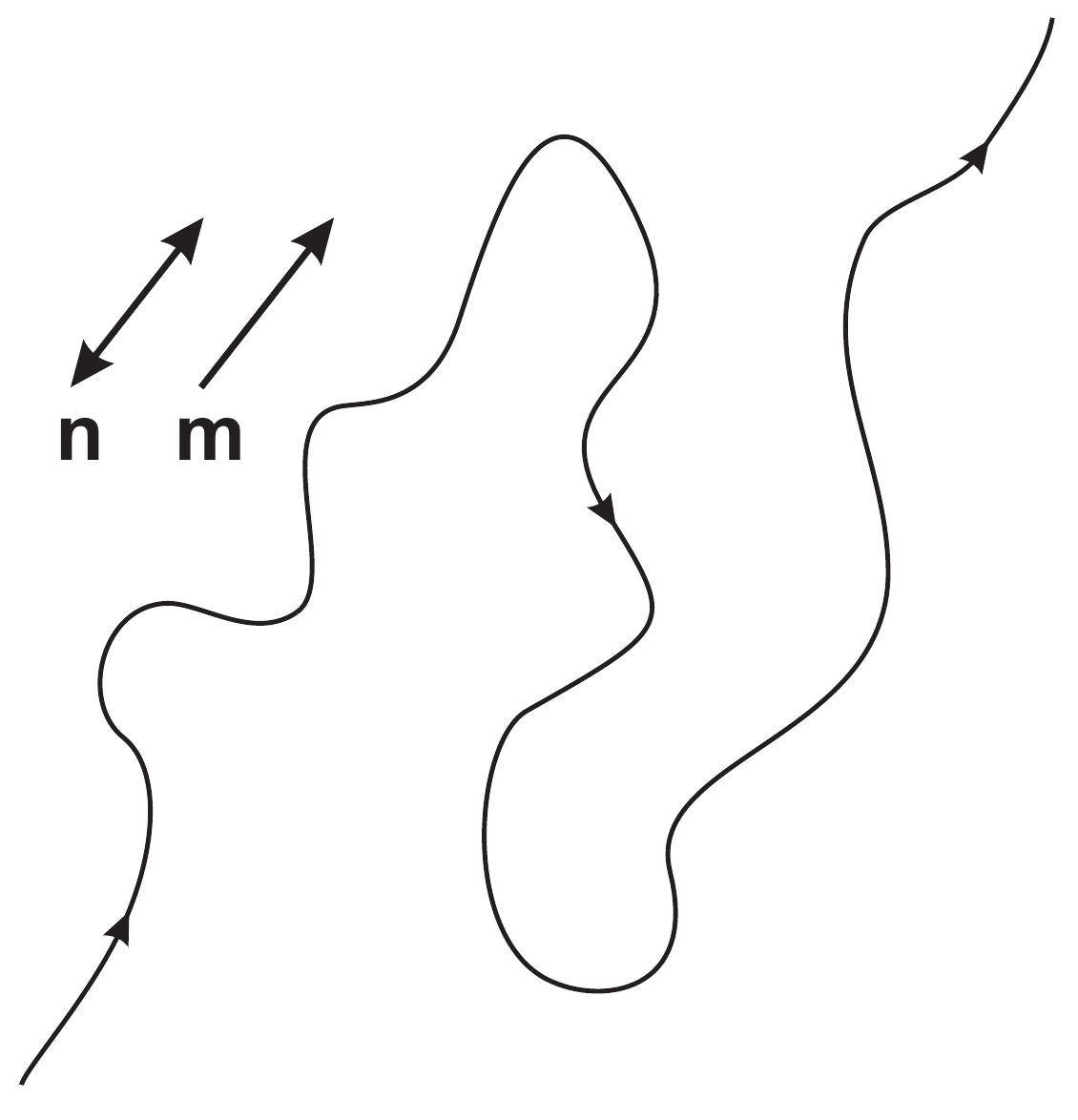}}\hspace{5mm}  
		\subfigure[]{\includegraphics[width=35mm]{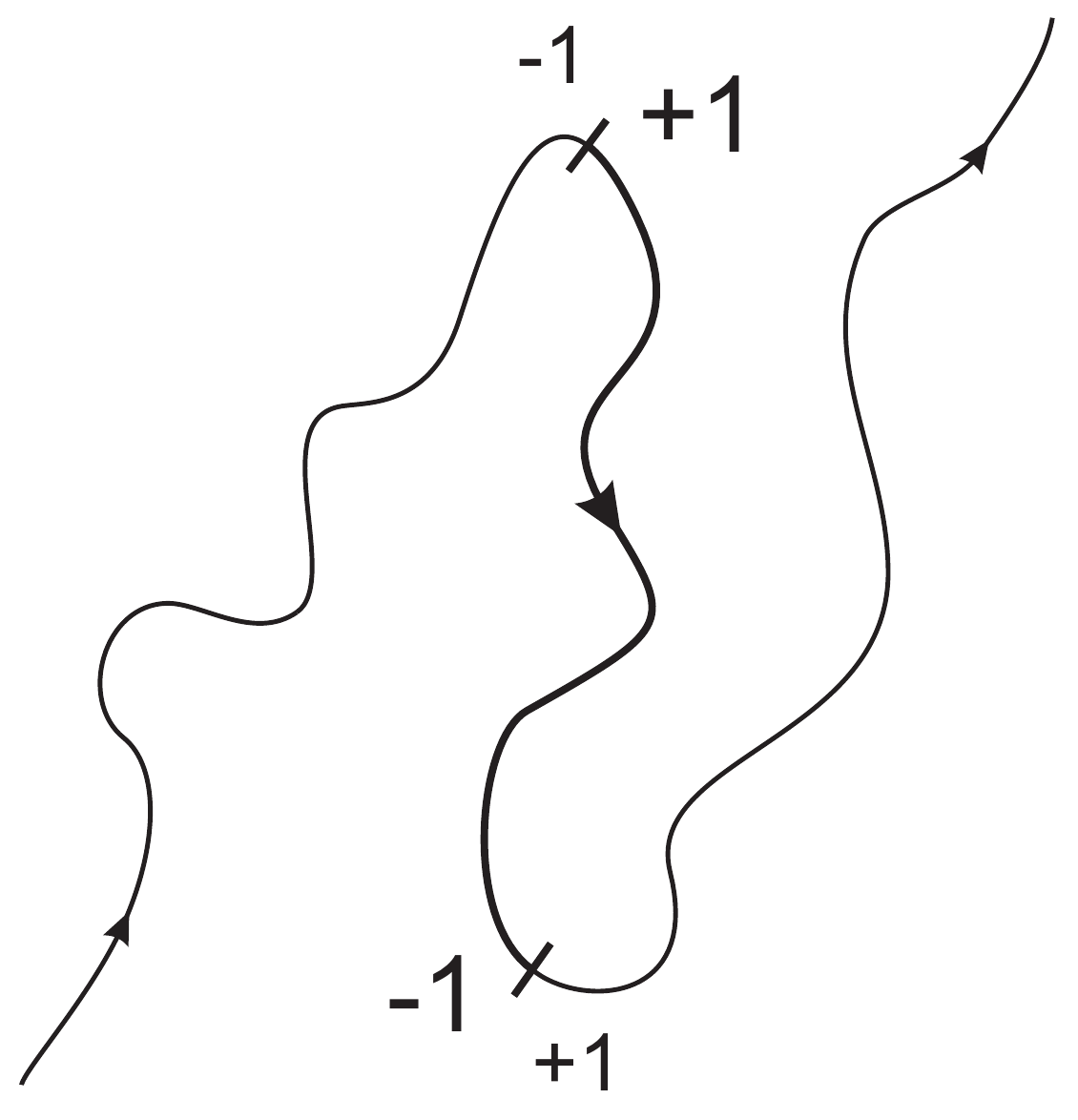}}\hspace{5mm}
		\subfigure[]{\includegraphics[width=35mm]{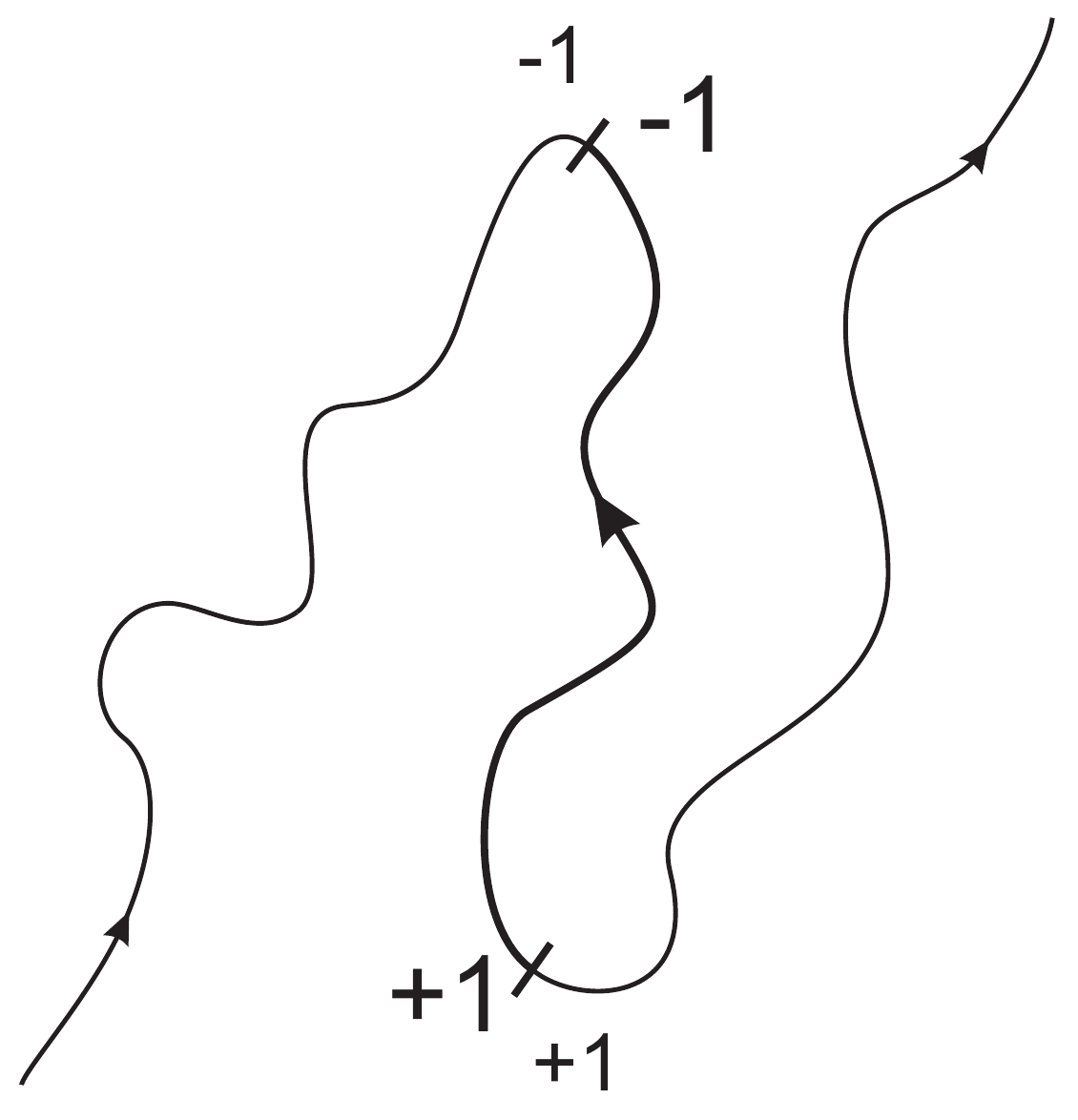}}
	}
	\caption{The purely formal procedure of cutting the chains at points of backfoldings and reversing the segments that are backfolded with respect to $\bf m$. One such segment is shown in (a). The cuts in (b) introduce a coinciding source and a sink each and have no effect. (c) By formally reversing the direction of the backfolded segment, a pair of separated $+2$ source and $-2$ sink is created, while the polar order of the system is increased.}
\label{fig:cutting}
\end{center}
\end{figure}

One can circumvent the inherent destruction of polar orientational order due to chain backfolding, by introducing virtual (imaginary) cuts in every chain as soon and each time it gets {\it backfolded} with respect to the nematic director, Fig.~\ref{fig:cutting}. Let us assign, arbitrarily but globally, an arrow to the director (in this context we are not worried by the resulting branch cuts in the case when topological disclinations are present) to get a {\it nematic vector} $\bf m$, Fig.~\ref{fig:cutting}a. The virtual cut is made whenever ${\bf t}(s)\cdot{\bf m}$ changes sign, Fig.~\ref{fig:cutting}b, which is also the definition of the chain backfolding (one should not confuse it with the hairpin). Every cut generates a $-1$ chain sink and a $+1$ chain source that coincide and thus add to zero. Such cuts are therefore purely imaginary and have no observable effect. However, now one can, without any physical change, reverse all the segments between those virtual cuts that have ${\bf t}(s)\cdot{\bf m}<0$, thereby also swapping the sources and sinks of these segments, Fig.~\ref{fig:cutting}c. Instead of the previous coinciding sources and sinks we now generated $+2$ sources and $-2$ sinks, separated in the directions upstream and downstream with respect to $\bf m$, respectively, and in particular, we created a nonvanishing macroscopic polar order $a=\vert{\bf a}\vert$ of chain tangents --- the {\it recovered polar order} ---, for which ${\bf a}\parallel{\bf m}$ holds due to the head-tail symmetry of nematic ordering. Note again that the system of chains was not modified physically in any way by this purely formal process.

Let us introduce a general source field $\rho^{\pm s}({\bf r})$ in the vectorial conservation Eq.~(\ref{vectorial}), now reading 
\begin{equation}
	\nabla\cdot{\bf j}=\rho^{\pm s}.
	\label{vectorial_rho_pms}
\end{equation}
For a uniform distribution of backfoldings that is expected in a nondeformed equilibrium configuration, the coarse-grained density of the new $\pm 2$ sources $\rho^{\pm s}=2\rho^{+2}-2\rho^{-2}$ is still zero. What is different, however, is the susceptibility $G$ of the $\rho^{\pm s}$ variations, given by the equilibrium densities of chain heads $\rho_0^{-2}$ and tails $\rho_0^{+2}$ in the  noninteracting ideal gas model \cite{nelson}, $G = k_B T/(\rho_0^{+2} + \rho_0^{-2})$.

Crucially, the recovered polar order {\it does not} depend on the density of backfoldings (for a given shape of the monomer orientational distribution function, its value is rigidly connected to the degree of nematic quadrupolar order), whereas the densities $\rho_0^{-2}$ and $\rho_0^{+2}$ {\it do} --- they are equal to the densities of the corresponding backfoldings. This, however, has physical implications described by the vectorial conservation law Eq.~(\ref{vectorial_rho_pms}) and the free energy of deviations of the source densities $\rho^+$, $\rho^-$, $\rho^{+2}$, $\rho^{-2}$
from their equilibrium values $\rho_0^+$, $\rho_0^-$, $\rho_0^{+2}$, $\rho_0^{-2}$, 
\begin{equation}
	f^{\pm s}(\rho^+-\rho_0^+,\rho^--\rho_0^-,\rho^{+2}-\rho_0^{+2},\rho^{-2}-\rho_0^{-2}).
	\label{f_pms_general}
\end{equation}
In other words, the introduction of the backfolding cuts $\rho_0^{+2}$ and $\rho_0^{-2}$ makes the $\rho^{\pm s}$ variations cheaper, thereby weakening the vectorial constraint for the recovered polar order. 

Let us reduce the unnecessary complexity of the sources and their free energy cost Eq.~(\ref{f_pms_general}) and show this explicitly.
The number of chain heads equals the number of chain tails and therefore
\begin{equation}
	\rho_0^+=\rho_0^-\equiv {\textstyle{1\over 2}}\rho_0^{\pm},
\end{equation}
the apolar symmetry of the nematic phase furthermore requires
\begin{eqnarray}
	\rho_0^{+2}=\rho_0^{-2} &\equiv& {\textstyle{1\over 2}}\rho_0^{\pm 2},
	\label{rho0pm2}\\
	\rho^{+}-\rho_0^{+}=-(\rho^{-}-\rho_0^{-}) &\equiv& {\textstyle{1\over 2}} \Delta\rho^\pm, \\
	\rho^{+2}-\rho_0^{+2}=-(\rho^{-2}-\rho_0^{-2}) &\equiv& {\textstyle{1\over 2}} \Delta\rho^{\pm 2}. 
\end{eqnarray}
To simplify life further, in a model we may describe both types of sources, i.e., chain ends $\Delta\rho^\pm$ and chain backfoldings $\Delta\rho^{\pm 2}$, on an equal basis, writing
$\rho_0^{\pm 2}=\alpha \rho_0^\pm$ and therefore also $\Delta\rho^{\pm 2}=\alpha \Delta\rho^\pm$, where $\alpha$ is a parameter of the system.
Hence, both types of sources can be expressed in terms of the total strength of sources
\begin{equation}
	\rho^{\pm s} = \Delta\rho^\pm + 2 \Delta\rho^{\pm 2}
	\label{sources_total}
\end{equation}
as
\begin{equation}
	\Delta\rho^\pm = {\rho^{\pm s}\over 1+2\alpha} \qquad {\rm and} \qquad 	\Delta\rho^{\pm 2} = {\alpha\rho^{\pm s}\over 1+2\alpha},	
\end{equation}
while the free energy cost of the sources Eq.~(\ref{f_pms_general}),
\begin{equation}
	f^{\pm s} = {1\over 2}{k_B T\over\rho_0^\pm}\left(\Delta\rho^\pm\right)^2 +
		    {1\over 2}{k_B T\over\rho_0^{\pm 2}}\left(\Delta\rho^{\pm 2}\right)^2
\end{equation}
is expressed as a single quadratic term,
\begin{equation}
	f^{\pm s} = {1\over 2} G \left(\rho^{\pm s}\right)^2,
	\label{f_pms}
\qquad {\rm with} \qquad
	G = {k_B T\over\rho_0^\pm} {1+\alpha\over(1+2\alpha)^2}.
\end{equation}
In the absence of the backfolding cuts, $\alpha = 0$ and Nelson's result \cite{nelson} for the susceptibility $G$ is recovered. As soon as the density of backfoldings increases, however, $G$ is lowered and eventually vanishes in the strong backfolding limit $\alpha\to\infty$.

To recapitulate, the recovered polar order is insensitive to the degree of chain backfolding and is thus a well-defined quantity. The strength of the vectorial constraint for the recovered polar order, however decreases with increasing degree of chain backfolding. Moreover, in the limit of strong backfolding, this constraint vanishes completely.
The strength of the vectorial constraint for the recovered polar order thus depends on the density of backfoldings (besides the density of chain heads/tails) and vice versa, in the continuum description the semi-flexibility of the polymer chain is controlled exactly and exclusively by the strength of this constraint.

In the light of this Section, the vectorial conservation law can thus be, perfectly rigorously, applied also to apolar nematic ordering. However, this is the conservation law for the recovered polar order and not for the quadrupolar order.
Actually, this reasoning must have been more or less tacitly assumed every time the vectorial constraint was applied on the nematic director. At the end of Section \ref{sec:s=1} it will become evident, why and when this simplistic standard application is incorrect.

\section{Connection between tensorial and vectorial conservation laws}

\noindent
It is illuminating to study the tensorial conservation law in its integral form. Being first order in gradients, the integrated constraint can be conveniently compared with the vectorial constraint.
Starting with the general tensorial conservation law Eq.~(\ref{tensorial}),
we can integrate it once to obtain
\begin{equation}
	\partial_j \left[\rho\ell_0\left(Q_{ij}+\textstyle{1\over 2}\delta_{ij}\right)\right] = 
		\textstyle{3\over 2} \left(g_i^+ - g_i^-\right) + {\textstyle{3\over 2}}\epsilon_{ijk}\partial_j \Psi_k,
		\label{curvature_mac}
\end{equation}
where the second term on the right-hand side is divergence-free.

To recognize the meaning of the divergence-free component, along the lines of ref.~\cite{svensek} we revert to the microscopic fields, writing
\begin{eqnarray}
	\partial_j \tilde{J}_{ij}^{mic}({\bf x}) &=& {\textstyle{3\over 2}}
	\int_{{\bf x}(s)} ds ~{dx_i(s)\over ds}{dx_j(s)\over ds} {\partial\over\partial x_j}\delta({\bf x} - {\bf x}(s)) = \dots\nonumber\\
	\dots& = & -{\textstyle{3\over 2}}\left[t_i(L)\delta({\bf x} - {\bf x}(L))-t_i(0)\delta({\bf x} - {\bf x}(0))\right] +
	{\textstyle{3\over 2}} \int_{{\bf x}(s)} ds ~{d^2 x_i(s)\over ds^2} \delta({\bf x} - {\bf x}(s)),
	\label{curvature_mic}
\end{eqnarray}
where the chain curvature
\begin{equation}
	{d^2 x_i(s)\over ds^2} \equiv 2 \delta(s-s^{h})N_i^{h} + \kappa^{ns}(s)N_i^{ns}(s)
\end{equation}
besides nonsingular ($\kappa^{ns}$) may have also singular contributions (in general these may be singular kinks of any angle, but here we restrict ourselves to hairpins only); a sum over all hairpins is omitted for brevity.
We thus have
\begin{equation}
	\partial_j \tilde{J}_{ij}^{mic}({\bf x}) = 
	{\textstyle{3\over 2}}\left\{
	t_i(0)\delta({\bf x} - {\bf x}(0))-t_i(L)\delta({\bf x} - {\bf x}(L))
	+ 2N_i^{h}\delta({\bf x} - {\bf x}(s^{h}))
	+ \int_{{\bf x}(s)} ds ~ \kappa^{ns}(s)N_i^{ns}(s)\delta({\bf x} - {\bf x}(s))
	\right\}.
	\label{curvature_mic_nice}
\end{equation}
From here it is already clear that point-like hairpins 
are analogous to chain beginnings/endings. 
We can rewrite Eq.~(\ref{curvature_mac}) by coarse graining 
Eq.~(\ref{curvature_mic_nice}):
\begin{equation}
	\partial_j \tilde{J}_{ij} = \partial_j \left[\rho\ell_0\left(Q_{ij}+\textstyle{1\over 2}\delta_{ij}\right)\right] = {\textstyle{3\over 2}} 
	\left(g_i^+ - g_i^- + 2 h_i + \rho\ell_0 k_i \right),
	\label{curvature_mac_nice}
\end{equation}
where $\bf h$ is now the density of hairpin principal normals, while
\begin{equation}
	k_i = {1\over L({\bf x})}\int_{{\bf x}(s)\in V({\bf x})} ds ~ \kappa^{ns}(s)N_i^{ns}(s)
\end{equation}
is the average nonsingular chain curvature field; $L({\bf x})$ is the total length of the polymer within the coarse-graining volume $V({\bf x})$ centered at ${\bf x}$.
Hence, we learn that
\begin{equation}
	\epsilon_{ijk}\partial_j \Psi_k = \rho\ell_0 k_i,
\end{equation}
while the contribution of the curvature singularities (point-like hairpins) --- which is generally not divergence-free --- has been exempt from $\bf k$ and added to the sources of the tensorial conservation law Eq.~(\ref{tensorial}), now reading
$\textstyle{3\over 2} \partial_i \left(g_i^+ - g_i^- + 2h_i\right)$.

We note that in the context of the tensorial conservation law point-like hairpins are only a formal abstraction --- in reality they need not exist, i.e., the chain tangent can be always regarded continuous, no matter how large the curvature; therefore ${\bf h}=0$, and Eq.~(\ref{curvature_mac_nice}) will serve merely to establish a link with the vectorial conservation (see below).
One might, however, have a specific physical reason to bring in also point-like hairpins and/or kinks: in some polymers such as DNA discrete kinks of the polymer chain might be present as a chemical configurational entity alternative to large-scale loops of distributed curvature \cite{Vologodskii}. In this case, Eq.~(\ref{curvature_mac_nice}) tells us how to account for such kinks explicitly in the conservation law ($\bf h$ 
is then a new variable that enters also the energy functional). The presence of thermalized kinks or even better, their generation by the addition of certain DNA binding proteins \cite{Tkachenko}, might constitute an interesting new way to actually observe and control the role of the tensorial continuity constraint.

\subsection{Vectorial conservation as a special case}
\label{sec:s=1}

\noindent
To establish the connection with the nematic director $\bf n$ (which is a natural reduction towards the vectorial order parameter), in Eq.~(\ref{curvature_mac_nice}) we use the uniaxial ansatz
\begin{equation}
	Q_{ij} = \textstyle{3\over 2}s\left(n_i n_j - \textstyle{1\over 3}\delta_{ij}\right)
	\label{uniaxial}
\end{equation}
to get
\begin{equation}
	\ell_0\left\{\textstyle{3\over 2}\left[n_j\partial_j(\rho s)\right]n_i + \textstyle{3\over 2}\rho s\left[n_i\partial_j n_j + n_j\partial_j n_i\right] + \textstyle{1\over 2}\partial_i\left[\rho(1-s)\right]\right\} =
	\textstyle{3\over 2} \left(g_i^+ - g_i^- + 2h_i + \rho\ell_0 k_i\right).
	\label{uniaxial_conservation}
\end{equation}

In the case of a perfect {\it local} orientational order, $s=1$, the chain end tangents ${\bf g}^\pm\parallel{\bf n}$ and the hairpin principal normals ${\bf h}\parallel{\bf n}$ are aligned with the director, whereas ${\bf k}\perp{\bf n}$ is perpendicular. In this case we have
\begin{equation}
	\ell_0\left\{{\textstyle{3\over 2}}\left[n_j\partial_j\rho + \rho\, \partial_j n_j\right]n_i 
	+ {\textstyle{3\over 2}}\rho\, n_j\partial_j n_i \right\} =
	\textstyle{3\over 2} \left(g_i^+ - g_i^- + 2h_i + \rho\ell_0 k_i\right)
\end{equation}
and hence in the directions parallel and perpendicular to $\bf n$, respectively:
\begin{eqnarray}
	\ell_0\left\{n_j\partial_j\rho + \rho\, \partial_j n_j\right\} &=& g^+ - g^- + 2h
	\label{parallel} \\
	n_j\partial_j n_i &=& k_i.
\end{eqnarray}
Thus, in the limit $s=1$ the vectorial conservation law is recovered exactly, with the hairpins acting as $\pm 2$ sources or sinks, while the macroscopic director bending field is equal to the local chain curvature.
We also see that in case chain backfoldings are present (due to $s=1$ these must be point-like hairpins), Eq.~(\ref{parallel}) is exactly the vectorial conservation law Eqs.~(\ref{vectorial_rho_pms}) and (\ref{sources_total}) for the recovered(!) polar order in the limit $a=1$ (the statement holds also in the absence of hairpins, obviously).

As soon as $s<1$, on the other hand, the directions $\parallel{\bf n}$ and $\perp{\bf n}$ are not decoupled in the conservation law Eq.~(\ref{uniaxial_conservation}), since $\partial_i\left[\rho(1-s)\right]$, $g_i^\pm$, $h_i$, and $k_i$ all point in general directions. In this case the integrated conservation law Eq.~(\ref{curvature_mac_nice}) is not useful and the original tensorial conservation law Eq.~(\ref{tensorial}) must be used. Since $s<1$ in all real systems, the tensorial conservation law Eq.~(\ref{tensorial}) cannot be in general reduced to the vectorial one and thus represents a new, distinct constraint. 
Only in the limit of perfect orientational order, $s=1$, the tensorial conservation law and the vectorial conservation law for the recovered polar order are indeed completely equivalent. Hence, it is only in this limit that the use of the vectorial conservation law for the nematic director (provided that the hairpins are included in the sources, of course) is correct. The stronger the deviation from perfect orientational order, the more inaccurate the vectorial conservation law description becomes, invalidating the results obtained from the assumption of its general validity.

\subsection{Compatibility of the conservation laws for $s<1$}

\noindent
In the context of no backfolding limit mentioned in Sec.~\ref{sec:intro}, it is important to address the relation between the tensorial and the vectorial conservation laws for the general case $s<1$. From what has been shown above, we understand that for $s<1$ the two conservation laws cannot be equivalent. However, in the limit of weak chain backfolding, both tensorial and vectorial (of the recovered polar order) constraints are effective in principle (the latter due to the small density of the backfolding cuts). Therefore the compatibility of the two constraints must be inspected to ensure that they are not in an incurable conflict, in particular for long chains without backfoldings, when both constraints are rigid.
This is certainly crucial for a consistent definition of a combined, generalized conservation law.


Retaining the uniaxial ansatz Eq.~(\ref{uniaxial}), for $\rho = {\rm const}$, $1>s={\rm const}$, and in the absence of sources the tensorial constraint Eq.~(\ref{tensorial}) reads
\begin{equation}
	\partial_i\left(n_i\partial_j n_j + n_j\partial_j n_i\right) = 0,
\end{equation}
which is incompatible with the no-splay requirement ($\partial_j n_j=0$) of the vectorial constraint, as in the second term it includes also the director bend deformation. In the following we show that this peculiarity of the bend deformation is connected with the arbitrarily imposed uniaxiality of the $\sf Q$-tensor.

Let us now allow a general form of the $\sf Q$-tensor,
\begin{equation}
	Q_{ij} = {\textstyle{3\over 2}}s\left(n_i n_j -{\textstyle{1\over 3}}\delta_{ij}\right)
		+{\textstyle{1\over 2}}p\left(e_i^1 e_j^1 - e_i^2 e_j^2\right),
\end{equation}
where $p$ is the biaxiality and $({\bf n},{\bf e}^1,{\bf e}^2)$ is an orthonormal triad. With this ansatz, the tensorial constraint Eq.~(\ref{tensorial}) in the absence of sources reads
\begin{eqnarray}
	\partial_i\left\{
	 n_j\partial_j\left[\rho(s+{\textstyle{1\over 2}})\right]n_i+{\textstyle{3\over 2}}\rho s(\partial_j n_j)\,n_i
	+ {\textstyle{3\over 2}}\rho s\, n_j(\partial_j n_i) 
	+ {\textstyle{1\over 2}}\delta_{ij}^\perp\partial_j\left[\rho(1-s)\right]\right. +
	\label{biax1}\\
	+\left.{\textstyle{1\over 2}}e_j^1\partial_j(\rho p)\,e_i^1-{\textstyle{1\over 2}}e_j^2\partial_j(\rho p)\,e_i^2
	+{\textstyle{1\over 2}}\rho p\,\partial_j\left(e_i^1 e_j^1 - e_i^2 e_j^2\right)
	\right\} = 0,
	\label{biax2}
\end{eqnarray}
where the gradients of $\rho$ and $s$ were split into longitudinal and transverse parts with respect to $\bf n$; 
$\delta_{ij}^\perp=\delta_{ij}-n_i n_j$.

Consider a uniaxial configuration that undergoes a bend deformation at constant $\rho$ and $s$, while at the same time the vectorial constraint rigidly requires $\nabla\cdot{\bf m}=0$ and thus also $\nabla\cdot{\bf n}=0$. By inspecting the directions of the terms in Eq.~(\ref{biax1})-(\ref{biax2}) inside the $\partial_i\{\}$ bracket, one can verify that the bend term ($\perp{\bf n}$) can be canceled by the first or second term in the line Eq.~(\ref{biax2}), provided that the gradient of the biaxiality becomes nonzero and the biaxial director ${\bf e}^1$ or ${\bf e}^2$ is parallel to the principal normal defined by the bend deformation. That the bend deformation is connected with the biaxiality is not surprising --- already in low-molecular-weight nematics the bend deformation, by asymmetry, induces a slight biaxiality. If the configuration is already biaxial, then also the last term of Eq.~(\ref{biax2}) is nonzero. In general, all three directions are involved, also $\rho$ and $s$ are varying, of course, and only $\partial_i\{\}$ must vanish rather than the expression inside the bracket, such that all degrees of freedom take their share in accommodating the tensorial constraint. 

It is thus indeed possible to satisfy both constraints simultaneously, even if they are rigid. Moreover, it is possible that the no-splay requirement does not affect the bend deformation, as in the case of the vectorial constraint alone. To what extent the tensorial constraint will be accommodated by variations of ${\bf n}$, $s$, $p$, ${\bf e}^1$, and $\rho$ is controlled by the relative free energy cost of these variations. The same holds for variations of $\bf a$ and $\rho$ connected by the vectorial constraint.

Finally, it is important to recognize the following point. Consider the case where the biaxiality is initially zero and the terms in the transverse directions vanish inside the $\partial_i\{\}$ bracket of Eq.~(\ref{biax1})-(\ref{biax2}) as before. The tensorial condition Eq.~(\ref{biax1})-(\ref{biax2}) is met if the longitudinal terms inside the bracket satisfy $n_j\partial_j\left[\rho(s+{\textstyle{1\over 2}})\right]+{\textstyle{3\over 2}}\rho s(\partial_j n_j) = 0$.
This requirement is very similar to the vectorial constraint $m_j\partial_j(\rho a)+\rho a(\partial_j m_j)=0$, where ${\bf a}=a{\bf m}$, but not identical. Even for $\rho={\rm const}$, unless $s=1$, the connection between splay and variations of the moduli is slightly different. This cannot be lifted --- it is due to the inherent difference between polar and quadrupolar moments of orientational order.

The reduction from the full $\sf Q$-tensor to the frequently used uniaxial director description is, in the context of the conservation laws, not an automatism as we have just seen. One can make various assumptions/models of how the omitted degrees of freedom entering the tensorial conservation behave when the reduction is made. For example, just using the uniaxial ansatz in Eq.~(\ref{tensorial}) forbids any biaxiality and necessarily introduces the direct connection between splay and bend in Eq.~(\ref{biax1}) that is not present in the case of the vectorial conservation law.


\section{Formulation of a generalized conservation law}

\noindent
The quadrupolar order can be described as usually by the $\sf Q$-tensor and a corresponding Landau-de Gennes nematic free energy functional $f_{n\rho}({\sf Q},\rho)$, which includes also the part that penalizes density variations. The conservation law is independent of the particular form of this functional, and we will hence not specify it further. The implementation of the full theory of the quadrupolar ordering thus assumes a concrete, non-universal form of this free energy functional and a universal implementation of the conservation law.

We first write the tensorial conservation law Eq.~(\ref{tensorial}) in terms of the $\sf Q$-tensor as,
\begin{equation}
	\partial_i\partial_j \left[\rho\ell_0\left(Q_{ij}+\textstyle{1\over 2}\delta_{ij}\right)\right] = 
		\textstyle{3\over 2} \partial_i g_i,
	\label{tensorial_conservation_1}
\end{equation}
where ${\bf g}({\bf r})$ is the density of chain end tangents pointing away from the end towards the chain. In general, ${\bf g}({\bf r})$ is an additional variable of the system (= 1 vector variable, since chain heads and tails are indistinguishable).
Moreover, at the same time the vectorial conservation law Eq.~(\ref{vectorial_rho_pms}) holds for the recovered polar order $\bf a$,
\begin{equation}
	\partial_i(\rho\ell_0 a_i) = \rho^{\pm s},
	\label{vectorial_conservation_1}
\end{equation}
where $\rho^{\pm s}$ is the total density of sources (chain ends and backfoldings) given in Eq.~(\ref{sources_total}).
As discussed in Sec.~\ref{sec:recovered_polar}, the recovered polar order ${\bf a}$ is assumed to be parallel to the nematic vector $\bf m$, while its magnitude $a = \langle\cos\theta\rangle$ is connected with the modulus of the quadrupolar nematic order $s=(3\langle\cos^2\theta\rangle-1)/2$,
\begin{equation}
	a^2 = {2s+1\over 3} A, \quad \langle\cos\theta\rangle^2 = A\langle\cos^2\theta\rangle,
	\label{moduli}
\end{equation}
where $A$ depends only on the orientational distribution function of the monomers and is assumed to be fixed.

Thus, the vectorial conservation Eq.~(\ref{vectorial_conservation_1}) for the recovered polar order is in fact an {\bf additional constraint for the director principal axis} of the ${\sf Q}$-tensor. In the case when the density of chain backfoldings $\rho_0^{\pm 2}$ of Eq.~(\ref{rho0pm2}) is high (small global persistence length), the free energy cost Eq.~(\ref{f_pms}) of the variation $\rho^{\pm s}$ is small. In this limit the vectorial constraint Eq.~(\ref{vectorial_conservation_1}) becomes ineffective and we are left with the tensorial conservation Eq.~(\ref{tensorial_conservation_1}) only. If, on the other hand, the density of the backfoldings is low, the additional constraint Eq.~(\ref{vectorial_conservation_1}) is strong and in this way implements the increased stiffness of the chains, which cannot be controlled by the tensorial conservation.

In general, the source fields $\nabla\cdot{\bf g}({\bf r})$, $\Delta\rho^\pm ({\bf r})$, and $\Delta\rho^{\pm 2} ({\bf r})$ are additional variables of the apolar polymer nematic system besides the ${\sf Q}({\bf r})$-tensor and the polymer density $\rho ({\bf r})$, for which one should write down additional couplings in the free energy functional. In a first approach, they can be however treated in a minimal spirit --- as a result of non-uniform distributions of the ideal gasses of chain ends and backfoldings with equilibrium densities $\rho_0^\pm$ and $\rho_0^{\pm 2}$, respectively. 
In this model, the free energy cost of the sources is purely entropic and is expressed in a form of Eq.~(\ref{f_pms}) or alike. A similar free energy contribution can be set up also for $\nabla\cdot{\bf g}$.
Within this simple model of the sources, the source densities do not appear explicitly as additional variables, since their quadratic free energy contributions can be expressed directly via the conservation laws Eqs.~(\ref{tensorial_conservation_1}) and (\ref{vectorial_conservation_1}) which now take the form of penalty potentials,
\begin{equation}
	f = f_{n\rho}({\sf Q}, \rho) 
	+ {\textstyle{1\over 2}} G \left[\partial_i (\rho\ell_0{a_i})\right]^2
	+ {\textstyle{1\over 2}} H \left\{\partial_i\partial_j\left[\rho\ell_0(Q_{ij}+{\textstyle{1\over 2}}\delta_{ij})\right]\right\}^2.
	\label{f_full}
\end{equation}
In this case the only additional parameters of the system are the equilibrium densities of chain ends $\rho_0^\pm$ and chain backfoldings $\rho_0^{\pm 2}$, which define the strengths $G$ and $H$ of the penalty potentials.

Finally, the system of equations is closed with two additional conditions and corresponding Lagrange-multiplier terms to be added to Eq.~(\ref{f_full}), keeping $\bf a$ parallel to the director principal axis of $\sf Q$ and its magnitude $a$ connected with the nematic modulus $s$ according to Eq.~(\ref{moduli}). Such conditions can be cumbersome if expressed analytically (in particular the former), but can be treated quite naturally in numerical approaches.

\section{Conclusion}

\noindent
The vectorial and tensorial conservation laws for main-chain polymer nematogens are not in general equivalent nor are they interchangeable.  In fact, using the vectorial conservation law for the usual quadrupolar nematic order as described in Section \ref{sec:recovered_polar} is incomplete and generally inaccurate, except in the limit of perfect orientational ordering. There exist in general two major reasons why the tensorial conservation law must be taken into account in addition to the vectorial conservation law for the recovered polar order:

\begin{enumerate}
	\item In the strong backfolding limit, the vectorial constraint for the recovered polar order vanishes as we have seen, thereby completely decoupling density variations and director deformations. Due to the tensorial conservation law, however, another, although looser (second order in the gradient) constraint remains also in the strong backfolding limit (vanishing global persistence length) of ideally flexible chains.
	\item Quadrupolar nematic order is described by the nematic $\sf Q$-tensor, i.e., it is captured by averaging a tensorial quantity, not a vectorial one as in the case of the recovered polar order. The tensorial description comprises more degrees of freedom, which is reflected among others also in the fact that the sources of the tensorial conservation law (the right-hand side of Eq.~(\ref{tensorial})) besides the volume density of chain ends comprise also the direction of their tangents. As we have shown, only in the limit of perfect orientational order (where the chains are perfectly aligned, all backfoldings are hairpins and there is no other directions involved) both conservation laws are equivalent.
\end{enumerate}

We propose that the generalized conservation law for semiflexible main-chain polymer nematics with usual quadrupolar (tensorial) nematic ordering should build on both, the vectorial as well as the tensorial conservation laws, implemented in the effective free energy via the penalty potentials. 
If the ordered phase is also polar, then the vectorial conservation law directly holds for its polar order. If, however, the quadrupolar order of this polar phase is included in the modeling (e.g. if a measurement probes the quadrupolar order of the polar phase rather than its polar order), the generalized conservation law for the quadrupolar order applies to it in principle, notwithstanding the concurrent existence of the polar order. Since in this case the backfolding is weak, there is little difference whether the actual or the recovered polar order enters the generalized conservation law for the quadrupolar order, but in principle it {\sl should be} the recovered polar order.

The present analysis of the two continuity constraints, identifying the extent of polymer chain backfolding as the determinant of their relative importance, also suggests a possible method to detect and/or manipulate their strength. Namely, in the case of DNA, where sharp local kinks can be induced by the addition of certain DNA binding proteins, an interesting new way to actually observe and control the role of the tensorial continuity constraint might be accessible by following the ordering of the chains in the solution and/or by monitoring the observable defects on adsorption of DNA kinking proteins. This path is feasible and worth pursuing on the level of simulations as well as experiments.

\begin{acknowledgments}
\noindent
Authors acknowledge the support of the Slovenian Research Agency (ARRS Grants No. J1-7435, N1-0019, and P1-0055). 
\end{acknowledgments}


\begin{thebibliography}{99}
\bibitem{errata} A factor of 3/2 is missing in ref.~\cite{svensek} on the right-hand sides of Eqs.~(18), (20), (22), (24), (25), (26), and (27).
\end{thebibliography}
\end{document}